\documentstyle[psfig]{mn}

\newif\ifAMStwofonts

\ifoldfss
  \ifCUPmtlplainloaded \else
    \NewTextAlphabet{textbfit} {cmbxti10} {}
    \NewTextAlphabet{textbfss} {cmssbx10} {}
    \NewMathAlphabet{mathbfit} {cmbxti10} {} 
    \NewMathAlphabet{mathbfss} {cmssbx10} {} 
  \fi
  \ifAMStwofonts
    \ifCUPmtlplainloaded \else
      \NewSymbolFont{upmath} {eurm10}
      \NewSymbolFont{AMSa} {msam10}
      \NewMathSymbol{\upi}     {0}{upmath}{19}
      \NewMathSymbol{\umu}     {0}{upmath}{16}
      \NewMathSymbol{\upartial}{0}{upmath}{40}
      \NewMathSymbol{\leqslant}{3}{AMSa}{36}
      \NewMathSymbol{\geqslant}{3}{AMSa}{3E}

    \fi
  \fi
\fi 

\ifnfssone
  \newmathalphabet{\mathit}
  \addtoversion{normal}{\mathit}{cmr}{m}{it}
  \addtoversion{bold}{\mathit}{cmr}{bx}{it}
  \newmathalphabet{\mathbfit} 
  \addtoversion{normal}{\mathbfit}{cmr}{bx}{it}
  \addtoversion{bold}{\mathbfit}{cmr}{bx}{it}
  \newmathalphabet{\mathbfss} 
  \addtoversion{normal}{\mathbfss}{cmss}{bx}{n}
  \addtoversion{bold}{\mathbfss}{cmss}{bx}{n}
  \ifAMStwofonts
    \ifCUPmtlplainloaded \else
      \UseAMStwoboldmath
      \makeatletter
      \new@mathgroup\upmath@group
      \define@mathgroup\mv@normal\upmath@group{eur}{m}{n}
      \define@mathgroup\mv@bold\upmath@group{eur}{b}{n}
      \edef\UPM{\hexnumber\upmath@group}
      \new@mathgroup\amsa@group
      \define@mathgroup\mv@normal\amsa@group{msa}{m}{n}
      \define@mathgroup\mv@bold\amsa@group{msa}{m}{n}
      \edef\AMSa{\hexnumber\amsa@group}
      \makeatother
      \mathchardef\upi="0\UPM19
      \mathchardef\umu="0\UPM16
      \mathchardef\upartial="0\UPM40
      \mathchardef\leqslant="3\AMSa36
      \mathchardef\geqslant="3\AMSa3E
    \fi
  \fi
\fi 

\ifnfsstwo
  \DeclareMathAlphabet{\mathbfit}{OT1}{cmr}{bx}{it}
  \SetMathAlphabet\mathbfit{bold}{OT1}{cmr}{bx}{it}
  \DeclareMathAlphabet{\mathbfss}{OT1}{cmss}{bx}{n}
  \SetMathAlphabet\mathbfss{bold}{OT1}{cmss}{bx}{n}
  \ifAMStwofonts
    \ifCUPmtlplainloaded \else
      \DeclareSymbolFont{UPM}{U}{eur}{m}{n}
      \SetSymbolFont{UPM}{bold}{U}{eur}{b}{n}
      \DeclareSymbolFont{AMSa}{U}{msa}{m}{n}
      \DeclareMathSymbol{\upi}{0}{UPM}{"19}
      \DeclareMathSymbol{\umu}{0}{UPM}{"16}
      \DeclareMathSymbol{\upartial}{0}{UPM}{"40}
      \DeclareMathSymbol{\leqslant}{3}{AMSa}{"36}
      \DeclareMathSymbol{\geqslant}{3}{AMSa}{"3E}
    \fi
  \fi
\fi 

\ifCUPmtlplainloaded \else
  \ifAMStwofonts \else 
    \def\upi{\pi}
    \def\umu{\mu}
    \def\upartial{\partial}
  \fi
\fi

\title{Numerical Simulations of Weak Lensing Measurements}
\author[D. J. Bacon et al.]
{David~J.~Bacon,$^1$\thanks{E-mail: djb@ast.cam.ac.uk}
Alexandre~Refregier$^1$, Douglas~Clowe$^{2}$, \& Richard~S.~Ellis$^{1,3}$\\
$^1$ Institute of Astronomy, Madingley Road, Cambridge CB3 OHA, UK \\
$^2$ Max-Planck-Institut f\"ur Astrophysik, 85740 Garching, Germany\\
$^3$ California Institute of Technology, Pasadena CA 91125, USA}

\date{Accepted ---. Received ---; in original form ---.}

\pagerange{\pageref{firstpage}--\pageref{lastpage}}
\pubyear{2000}

\begin{document}

\maketitle

\label{firstpage}

\begin{abstract}
Weak gravitational lensing induces distortions on the images of
background galaxies, and thus provides a direct measure of mass
fluctuations in the universe. The distortion signature from
large-scale structure has recently been detected by several groups for
the first time, opening promising prospects for the near future.
Since the distortions induced by lensing on the images of background
galaxies are only of a few percent, a reliable measurement demands
very accurate galaxy shape estimation and a careful treatment of
systematic effects. Here, we present a study of a shear measurement
method using detailed simulations of artificial images. The images are
produced using realisations of a galaxy ensemble drawn from the HST
Groth strip. We consider realistic observational effects including
atmospheric seeing, PSF anisotropy and pixelisation, incorporated in a
manner to reproduce actual observations with the William Herschel
Telescope.  By applying an artificial shear to the simulated images,
we test the shear measurement method proposed by Kaiser, Squires \&
Broadhurst (1995, KSB). Overall, we find the KSB method to be reliable
with the following provisos. First, although the recovered shear is
linearly related to the input shear, we find a coefficient of
proportionality of about 0.8. In addition, we find a residual
anti-correlation between the PSF ellipticity and the corrected
ellipticities of faint galaxies. To guide future weak lensing surveys,
we study how seeing size, exposure time and pixelisation affect the
sensitivity to shear. We find that worsened seeing linearly increases
the noise in the shear estimate, while the sensitivity depends only
weakly on exposure time. The noise is dramatically increased if the
pixel scale is larger than that of the seeing. In addition, we study
the impact of overlapping isophotes of neighboring galaxies, and find
that this effect can produce spurious lensing signals on small
scales. We discuss the prospects of using the KSB method for future,
more sensitive, surveys. Numerical simulations of this kind are a
required component of present and future analyses of weak lensing
surveys.
\end{abstract}


\section{Introduction}
\label{intro}
Weak lensing provides a unique method to directly measure the mass
fluctuations on large scales in the universe (see Mellier 1999; Kaiser
1999; Bartelmann \& Schneider 1999 for recent reviews). This method
relies on the measurement of small, coherent distortions produced by
lensing upon the shapes of background galaxies. This effect is now
routinely used to map the mass of clusters of galaxies (see reviews by
Fort \& Mellier 1994, Schneider 1996). Recently, the technique was
extended to the field by several groups who reported the statistical
detection of weak lensing by large-scale structure (Wittman et
al. 2000; van Waerbeke et al. 2000; Bacon, Refregier \& Ellis 2000 (BRE);
Kaiser, Wilson \& Luppino 2000). More precise measurements of this
``cosmic shear'' from upcoming observations will provide invaluable
cosmological information (eg. Kaiser 1992; Jain \& Seljak 1997;
Kamionkowski et al.  1997; Kaiser 1998; Hu \& Tegmark 1998; van
Waerbeke et al. 1998).

Because the distortions induced by lensing are only of the order of
1\%, these measurements are very challenging. In particular, they
require tight control of systematic effects and a precise method for
the measurement of the shear. One of the potential weaknesses of the
cosmic shear programme is the step leading from the measurement of the
shapes of galaxies to the estimation of the lensing shear, in the
presence of an anisotropic PSF. The first method proposed to treat
this problem was that by Bonnet \& Mellier (1995).  A more general, and
now widely used, method was proposed by Kaiser, Squires \& Broadhurst
(1995, KSB) and further developed by Luppino \& Kaiser (1997) and
Hoekstra et al. (1998). Variations and alternatives to the KSB method
have since been presented by Kaiser (1999b), Rhodes et al. (1999) and
Kuijken (1999).

In this paper, we address the ellipticity-to-shear problem using
numerical simulations of artificial images. The numerical simulations
which have been used for this purpose in the past have been derived
either from HST images, degraded to match ground-based observations
(eg. KSB, Wittman et al. 2000), or by using ab-initio artificial galaxy
catalogs (eg. Kaiser 1999b). The former approach provides accurate
shape statistics for the simulated galaxies, but can only produce a
small simulated area. The latter approach allows the simulation of
arbitrarily large areas, but is not necessarily as realistic. Because
we are aiming at the demanding cosmic shear regime, we thus use a
hybrid method in which large realisations of artificial galaxy
images are drawn to reproduce the statistics of existing HST surveys.

Because it is widely used and more documented, we focus on the KSB
method, and test its reliability in realistic observational
conditions. For definitiveness, we consider the weak-lensing survey of
BRE, who used ground based observations with the William Herschel
Telescope (WHT).  We produce artificial galaxy catalogs generated from
random realisations based on the HST Groth Strip (Groth et al. 1994,
Rhodes 1999). By applying artificial shears to the simulated images,
we test both the systematic and statistical uncertainties of the
method. We also investigate how the shear signal degrades as a
function of seeing, exposure time and pixel size. This is of
considerable practical interest for the design of future weak lensing
surveys. In addition, we examine the impact of overlapping isophotes
on the shear signal, an effect which can potentially limit weak
lensing measurements on small scales. An independent study of the shear
measurement method using numerical simulations is presented in Erben
et al. (2000)

The paper is organised as follows. In \S\ref{cat}, we describe how we
generate the artificial galaxy and star catalogues. In \S\ref{im}, we
show how these are used to produced realistic images, which are
compared to observed WHT images. This is followed in \S\ref{KSB} be a
brief description of our implementation of the KSB method. In
\S\ref{results}, we present our results; the accuracy of
recovery of shear with the KSB method is discussed, and the occurrence
of an anti-correlation of shear with star ellipticity at low level is
noted. We demonstrate the degradation of signal-to-noise with
increasing seeing, exposure time and pixel size, and also discuss the
level at which overlapping isophotes will enhance the cosmological
shear signal. The results are discussed and summarized in
\S\ref{conclusion}.

\section[]{Simulated object catalogue}
\label{cat}

The first step in these simulations is to construct an object
catalogue.  To do so, we used the image statistics from the Groth
Strip, a deep survey taken with the Hubble Space Telescope (Groth et
al. 1994, Rhodes et al 1999). This HST survey is sampled at 0.1 arcsec
and thus effectively gives us the unsmeared (i.e. before convolution
with ground-level seeing) ellipticities and diameters of an ensemble
of galaxies suitable for simulations of ground-based observations. The
survey consists of a set of 28 contiguous pointings in $V$ and $I$,
with an area of approx. 108 arcmin$^2$ in a 3'.5 $\times$ 44'.0
region. The magnitude limit is $I \simeq 26$ (WFPC2 I band, F814W),
and the strip includes about 10,000 galaxies.

We use a SExtractor (Bertin \& Arnoults 1996) catalogue derived from
the entire strip by Ebbels (1998). It contains, for each object, a
magnitude determined by aperture photometry, and a diameter and
ellipticity derived from second-order moments using a top-hat weight
function. Armed with this catalogue, we model the multi-dimensional
probability distribution of galaxy properties (ellipticity - magnitude
- diameter) sampled by this catalogue. We find that the differential
galaxy counts as a function of $I$-magnitude are well described by
$\frac{dn}{dI} \simeq 10^{(a_0 + a_1 I + a_2 I^2)}$ galaxies
deg$^{-2}$ mag$^{-1}$, with $a_0=-19.0$, $a_1$=1.64 and $a_2=-0.027$;
the radius distribution is modelled as $r=0.095 \times 10^{(b_1+b_2 I
+ b_3)}$ arcsecs with $b_2=-0.14$ and with $b_1$ a gaussian
distribution with mean 3.75 and rms 0.098. Ellipticity components
$e_1$ and $e_2$ are described by gaussian probability distributions of
rms 0.34, and the position angle is randomly chosen. We draw from this
modeled distribution a catalogue of galaxies statistically identical
to the Groth strip distribution by Monte Carlo selection.

For definiteness, we aim to reproduce the conditions of our weak
lensing survey derived from observations with WHT (see BRE).  This
survey consisted of 14 independent blank fields observed the WHT prime
focus CCD Camera (field of view 8' $\times$ 16', pixel size 0.237'',
EEV CCD) in the $R$ band. A relevant issue is the number of stars
obtained: by tuning the galactic latitude ($30^\circ < b < 70^\circ$),
we required the fields to contain $\simeq 200$ stars with $R<22$ in
order to map carefully the PSF and the camera distortion across the
field of view. The integration time on these fields was 1 hour,
affording a magnitude limit of $R$=25.2 (all $R$ magnitudes quoted as
Vega magnitudes).

Given that the Groth strip is in $I$ while our data is in $R$, we
allowed a slight increase (multiplication by 1.2) to radius with
magnitude to better model the WHT images; this factor has been
included in the described model above. The number density-magnitude
dependence was found to fit very well without alteration.

We spatially distribute the galaxies with a uniform probability across
the field of view. Since the Groth strip does not contain enough stars
to create a good model, star counts with $R$-magnitude are modeled as
a power law ($\frac{dn}{dR} \propto R^{0.2}$) which is found to be a
good fit to the WHT data.

We assign a morphological class (elliptical, spiral or irregular) for
future image realisation to each galaxy using the results of Abraham
et al (1996). Specifically, we use their measured fraction for each
class as a function of magnitude.  After application of all of the
above procedure, we obtain our unlensed object catalogues.

To produce the lensed object catalogue, we sheared the galaxies in
the catalogue by calculating the change in the object ellipticity due
to lensing. This was done using the relation (Rhodes et al 1999):
\begin{equation}
\label{eq:e_gamma}
e'_i = e_i + 2(\delta_{ij} - e_i e_j)\gamma_j
\end{equation}
Since we are primarily interested only in the mean shear measured on
the field, we chose the imposed shear to be uniform over a given
field.

Stellar ellipticities (simulating tracking errors, atmospheric
effects, etc) are similarly chosen as uniform over a given field. The
ellipticity for each field is taken from a Gaussian probability
distribution with a standard deviation of
$\sigma_e^{*}=0.08$. This is conservatively chosen to be
slightly worse than the rms stellar ellipticity of the stars in our
WHT survey, for which $\sigma_e^{*} \simeq 0.07$.

\section[]{Image Realisation}
\label{im}

We create the artificial images using the IRAF {\tt Artdata}
package. This takes the star and galaxy catalogues, and plots the
objects with specified positions, ellipticity, magnitude,
diameter and morphology. Only exponential discs and de Vaucouleurs
profiles are supported. We model ellipticals and irregulars as de
Vaucouleurs profiles, and spirals as exponential discs.

Each pixel in these simulations is subdivided into a $10\times 10$
grid of subpixels. The appropriate subpixel flux for a star or galaxy
is calculated from the analytical intensity profile, and the
PSF convolution is similarly performed at the subpixel level.

We use the package to recreate several WHT-specific details: the
magnitude zero point is chosen to match the telescope throughput, the
stars and galaxies are convolved with the chosen elliptical PSF
(seeing chosen to be 0.8'' unless otherwise specified, ellipticity
dispersion 0.08), the image is appropriately pixellised (0.237'' per
pixel), Poisson photon noise for objects and sky background are
included, and gaussian CCD read noise (3.9 electrons) are added. The
appropriate gain (1.45 electrons / ADU) is included, and an
appropriate sky background (10.7ADU per sec) is imposed. The PSF
profile chosen is the Moffat profile, $I(r) = (1 + (2^{1/\beta}-1)
(r/r_{\rm scale})^2)^{-\beta}$, where $\beta$ = 2.5 and $r_{\rm
scale}$ is the seeing radius. The generalised radius $r$ is the
distance from the centroid, transformed so that the profile is
elliptical. This profile has wings which fall off more slowly than for
a gaussian profile, and provides a good description of our
seeing-dominated PSF.

\begin{figure}
\psfig{figure=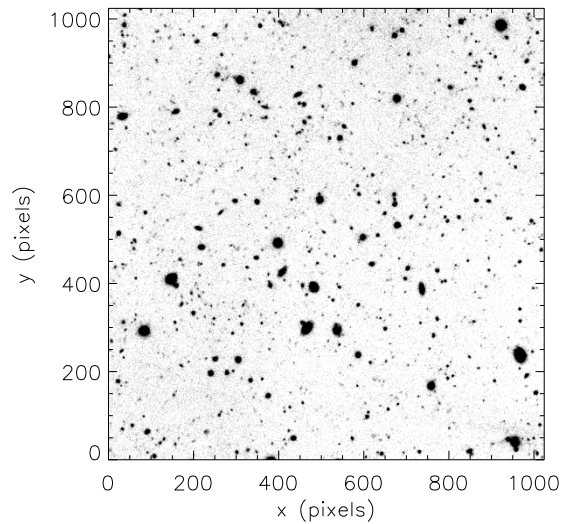,width=80mm} 
\caption{Detail of a real data image (WHT3). The area displayed is
4'$\times$ 4', while the full image is $8' \times 16'$.}
\label{fig:cirsi2}
\end{figure}

\begin{figure}
\psfig{figure=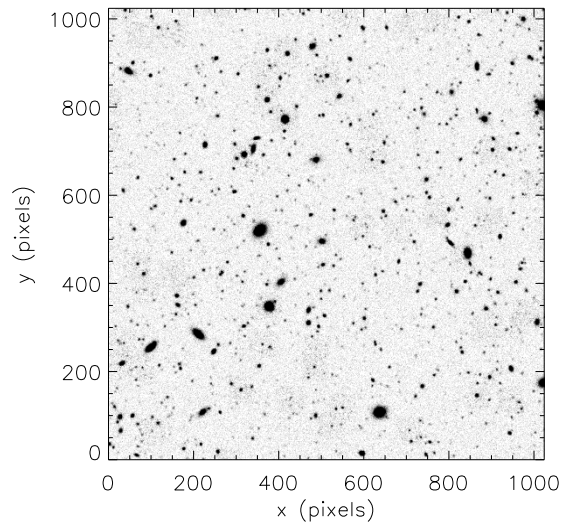,width=80mm} 
\caption{Detail (4'$\times$ 4') of a simulated image.}
\label{fig:simimage}
\end{figure}

An example 4' $\times$ 4' portion of a $16 \times 8'$ simulated field
is shown in figure \ref{fig:simimage}. It can be compared to an
observed WHT image shown in figure \ref{fig:cirsi2}. A global
impression can be obtained from the full $16 \times 8'$ fields plotted
in BRE. The simulated image lacks saturated stars ($R \la 18$); these
are, by construction absent from the simulated catalogue (see
size-magnitude figure in BRE), since such stars cannot be used in our
weak lensing analysis, and would be immediately excised if
present. The galaxy images appear to be very similar in the
simulations as compared with the data.

To compare these images quantitatively, we derived a measured object
catalog using the {\tt Imcat} routines (see \S\ref{KSB} below for
details). They provide the position, magnitude, half-light radius,
ellipticity and polarisability tensors for each object detected on the
image. 

\begin{figure}
\psfig{figure=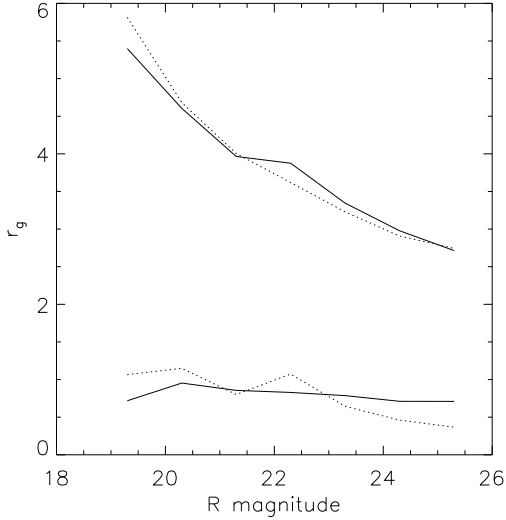,width=80mm} 
\caption{Galaxy size statistics with $R$ magnitude. The upper curves show 
the mean filter radius $r_g$ in unit magnitude bins; solid line shows
real data, dotted line shows simulated data. The lower curves show the
standard deviation on the radius in unit magnitude bins.}
\label{fig:magrg}
\end{figure}

The resulting distribution on the radius-magnitude plane is shown in
section 7 of BRE for both a simulated field and the observed WHT
field; the running mean and standard deviation of $r_g$ with magnitude
is shown on figure \ref{fig:magrg}. The distributions are similar,
with mean radius agreeing to within 0.2 pixels throughout the
magnitude range. The rms scatter in radius is somewhat larger at faint
magnitudes for the real data, due to the response of the measurement
software to the simple smooth profiles used in the simulations.

\begin{figure}
\psfig{figure=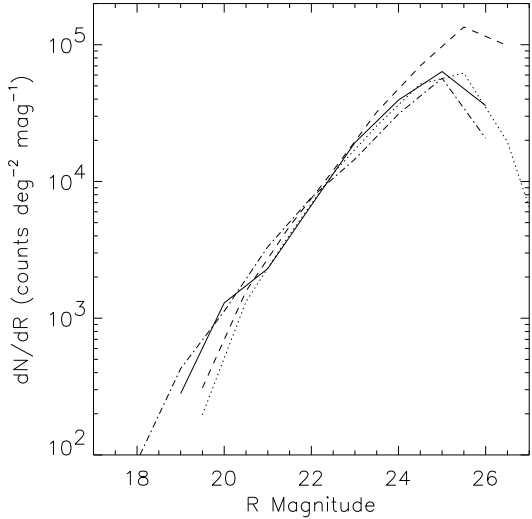,width=80mm} 
\caption{Differential galaxy number counts with $R$-magnitude (except
for real Groth data in $I$). Real WHT data (WHT3) is shown as solid
line; Groth strip number counts are shown as dash-dotted. The counts
for the simulated catalogue are shown dashed, and those recovered by
KSB from the simulated images is shown dotted.}
\label{fig:n}
\end{figure}

Figure \ref{fig:n} compares differential number counts with
$R$-magnitude for the simulations and real data. The counts derived
from the simulated image and from our real $R$-band fields agree very
well; the simulated image counts also agree well with the simulated
catalogue counts, with an unsurprising turn-over near the expected
magnitude completeness limit for the 1 hour exposure time ($R\la
25$). The simulations' close impersonation of real data number counts
is of importance for obtaining realistic results for, eg, overlapping
isophotes in section \ref{results}.

\begin{figure}
\psfig{figure=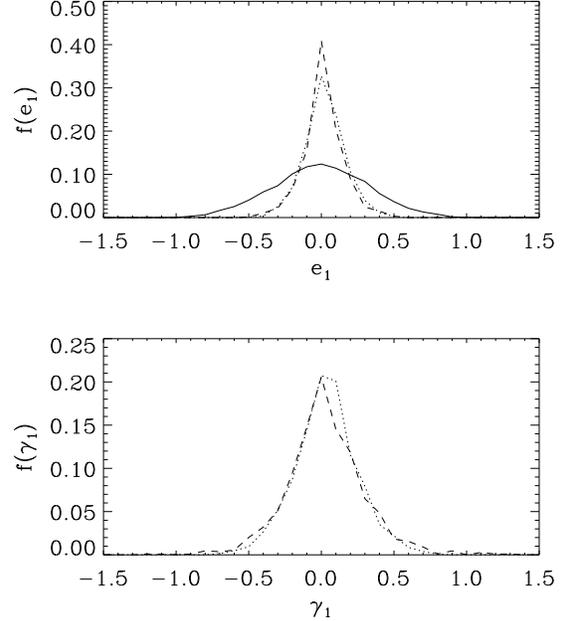,width=80mm}
\caption{Comparison of shape measures. The top panel shows the
normalised ellipticity $e_1$ distribution for initial
unsmeared simulated catalogue (solid) and smeared simulated (dotted)
and real (dashed) objects. The bottom panel shows the distribution of
the shear estimators for corrected simulated (dotted) and real
(dashed) objects.}
\label{fig:e}
\end{figure}

Figure~\ref{fig:e}a shows the ellipticity ($e_{1}$) distributions
$f(e_{1})$ for the {\it initial} simulated catalogue, and the {\it
smeared} simulated and real objects. The distributions were
normalised so that $\int de_1~f(e_1) \equiv 1$. As expected, smearing
reduces the ellipticity dispersion. The simulated and real smeared
distributions are remarkably similar. See section \ref{results} for
further discussion on this point.

\section[]{Shear measurement method}

\label{KSB}

Our shear measurement method is a version of the KSB method and was
described in detail in BRE. Here we summarize it and then describe
the relevant details of our specific implementation.

\subsection{Overview of the KSB method}

The KSB method derives the shear from the ellipticity of galaxies,
after correcting for the smearing by the PSF. We use the
implementation of the KSB method achieved by the {\tt imcat} software
kindly provided to us by Nick Kaiser.  

The routine {\tt hfindpeaks} first finds objects in each field by
convolving the image with smoothing kernels of different sizes. The
object radius $r_{g}$ is defined by the size of the kernel which
maximizes the signal-to-noise $\nu$ of the object. The routine {\tt getshapes}
then takes $r_{g}$ as the size of the gaussian weight function used to
measure quadrupole moments $I_{ij}$ of the object about its center of
light. The ellipticity of the object is then defined as $e_{i} \equiv
\{I_{11}-I_{22}, 2 I_{12}\} / (I_{11}+I_{22})$.

The next step in the KSB algorithm is to correct for the anisotropy
of the PSF. The corrected ellipticity of a galaxy $e^g_{\rm
corrected}$ is related to the observed smeared ellipticity $e^g_{\rm
smeared}$ by
\begin{equation}
e^g_{\rm corrected} = e^g_{\rm smeared} - P^g_{sm} p,
\label{eq:ecorrect}
\end{equation}
where the ellipticities are understood to denote the relevant
2-component spinor $e_i$, and $p$ is a measure of PSF anisotropy. The
tensor $P^g_{sm}$ is the smear polarisability, a $2\times 2$
matrix with components involving higher moments of surface
brightness. Since for stars $e^*_{\rm corrected}=0$, $p$ can be
measured using
\begin{equation}
p = (P^*_{sm})^{-1} e^*_{\rm smeared}.
\end{equation}

The lensing shear takes effect before the circular smearing of the
PSF. Luppino and Kaiser (1997) showed that the
{\it pre}-smear shear $\gamma$ averaged over a field can be
recovered using
\begin{equation}
\label{eq:gamma_est}
\gamma  = P_{\gamma}^{-1} {e^g_{\rm corrected}}
\end{equation}
where
\begin{equation}
\label{eq:p_gamma}
P_{\gamma} = P_{sh}^g - \frac{P_{sh}^*}{P_{sm}^*} P_{sm}^g.
\end{equation}
Here, $P_{sh}^g$ is the shear polarisability tensor for the galaxy
involving other higher order moments of the galaxy image.  The
quantities $P_{sh}^*$ and $P_{sm}^*$ are the shear and smear
polarisabilities calculated for a star interpolated to the position of
the galaxy in question. With the smear and shear polarisabilities
calculated by {\tt imcat}, we can therefore find an estimator for the
mean shear in a given cell.

\subsection{Specific Implementation}
\label{specific}
Firstly, we need to remove noisy detections. We applied a size limit
$r_g > 1.0$ to reject extraneous detections of very small objects
claimed by {\tt imcat}. We also applied a signal-to-noise $\nu > 15.0$
limit (see \S\ref{ani_test} for justification of this apparently very
conservative cut). To reduce the noise in our measurement, we also
remove highly elliptical objects with $e > 0.5$.

Stars were identified using the non-saturated stellar locus on the
magnitude--$r_h$ plane (see figure 11 in BRE), typically with $R
\simeq$ 19-22. In the data, the stellar ellipticity is a smooth
function of position on the field. We thus adopted an iterative
interpolation scheme to model this variation.
Specifically, we first fitted a 2-D cubic to the measured stellar
ellipticities, plotted the residual ellipticities $e^{res} = e^* -
e^{fit}$ and re-fitted after the removal of extreme outliers (caused
by galaxy contamination, blended images and noise). The stellar
ellipticity was kept constant in the simulations, but we nevertheless
fit the 2-D cubic for correction, as a means of retaining potential
systematic effects induced in this step.

In order to correct galaxies for anisotropic smear, we not
only need the fitted stellar ellipticity field, but also the four
component stellar smear and shear polarisabilities as a function of
position. Here a 2-D cubic is fit for each component of $P_{sm}^*$
and $P_{sh}^*$. Galaxies are then chosen from the magnitude-$r_h$
diagram by removing the stellar locus and objects with $\nu<15$,
$r_{g}<1$, $e>0.5$, as described above. From our fitted stellar
models, we then calculate $e^*$, $P_{sm}^*$ and $P_{sh}^*$ at each
galaxy position, and correct the galaxies for the anisotropic PSF
using equation (\ref{eq:ecorrect}). As a result, we obtain
$e^{g}_{{\rm corrected}}$ for all selected galaxies in each cell.

We then calculate $P_{\gamma}$ for the galaxies. We opt to treat
$P_{sh}^*$ and $P_{sm}^*$ as scalars equal to half the trace of
the respective matrices. This is allowable, since the non-diagonal
elements are small and the diagonal elements are equal within the
measurement noise (typical $P_{sm,11,22}^*$ = 0.10,
$P_{sm,12,21}^* < 5\times 10^{-4}$, $P_{sh,00,11}^*$ = 1.1,
$P_{sh,12,21}^* < 0.01$). 

With this simplification, we calculate $P_{\gamma}$ according to
equation~(\ref{eq:p_gamma}). $P_{\gamma}$ is typically a noisy
quantity, so we fit it as a function of $r_g$. We choose to treat
$P_{\gamma}$ as a scalar, since the information it carries is
primarily a correction for the size of a given galaxy, regardless of
its ellipticity or orientation. We thus plot $P_{\gamma}^{11}$ and
$P_{\gamma}^{22}$ together against $r_g$, and fit a cubic to the
combined points. Moreover, since $P_{\gamma}$ is unreliable for
objects with $r_g$ measured to be less that $r_g^*$, we remove all
such objects from our prospective galaxy catalogue. Finally, we
calculate a shear measure for each galaxy as in
Equation~(\ref{eq:gamma_est}), where the $P_{\gamma}$ is the fitted
value for the galaxy in question.

Because of pixel noise, a few galaxies yield extreme, unphysical,
shears $\gamma$. To prevent these from unnecessarily dominating the
analysis, we have removed galaxies with $\gamma>2$. 

This entire procedure provides us with an estimator of the shear
$\gamma$ for each galaxy. We can also calculate the mean shear
$\bar{\gamma}=\langle \gamma \rangle$ in a cell and its associated
error $\sigma[\bar{\gamma}]=\sigma[\gamma]/\sqrt{N}$, where $N$ is the
number of galaxies in a cell.

\section[]{Results}
\label{results}

\subsection{Ellipticity Distribution}
\label{e_dist}
We first compare the distribution of ellipticities and shear
estimators within a field. As we noted above, the uncorrected
ellipticity distribution of the simulated objects is very similar to
that of the data (Figure~\ref{fig:e}a). The distribution of the shear
estimators $\gamma_{1}$ after all corrections is shown on
Figure~\ref{fig:e}b, for each case. The agreement is good,
showing that the simulations faithfully reproduce the shape statistics
of the data, the central concern for weak lensing.

In general, the variance of the shear estimators results from several
effects: the intrinsic ellipticity dispersion, pixelisation, and pixel
noise. The latter effect is enhanced by the correction for the
isotropic smearing. The ellipiticity and shear rms dispersion,
$\sigma_{e} \equiv \langle e^2 \rangle^{\frac{1}{2}}$ and
$\sigma_{\gamma} \equiv \langle \gamma^2 \rangle^{\frac{1}{2}}$, at
different stages of the correction algorithm are listed in
Table~\ref{tab:variance}. The ellipticity dispersion observed in the
simulated image is reduced by the PSF smearing as compared to the
input ellipticity dispersion. (Pixel noise and pixelisation tends to
increase the observed dispersion, but the smearing dominates).  The
smearing is corrected for by the KSB method, leading to a re-increased
dispersion $\sigma_{\gamma}$ in the corrected shear estimator.

\begin{table*}
\centering \begin{minipage}{180mm}
\caption{Ellipticity and Shear rms dispersion at different stages
pf the correction method}
\label{tab:variance}
\begin{tabular}{lll} 
\hline dispersion & simulation & data \\
\hline 
$\sigma_{e}$ (input)     & 0.47 &  \\
$\sigma_{e}$ (measured)  & 0.20 & 0.20 \\
$\sigma_{\gamma}$ (final)& 0.31 & 0.39 \\  
\hline
\end{tabular}\\
\end{minipage}
\end{table*}

We can obtain an estimate of the relative contribution of pixel noise
and intrinsic dispersion using these results. In the absence of
weighting and smearing, the ellipticity is related to the shear by
$\epsilon = g \gamma$, where $g=(2-\langle \epsilon^{2} \rangle)$ (see
Eq.~[\ref{eq:e_gamma}] and Rhodes et al. 2000). For the simulations,
the input ellipticity dispersion is $\sigma_{e} \simeq 0.47$, yielding
$g \simeq 1.8$. As a result, the shear rms produced by the intrinsic
dispersion alone is $\sigma_{\gamma}^{\rm intrinsic} \approx
\sigma_{e}^{\rm input}/g \simeq 0.26$. The fact that this value is
close to the total shear disperion $\sigma_{\gamma} \simeq 0.31$
observed in the simulations shows that the intrinsic dispersion is
larger but comparable to that produced by pixel noise and
pixelisation. This considerations should be kept in mind in planning
the exposure time of weak lensing surveys. We will study the impact of
worsened seeing and larger pixel sizes in \S\ref{seeing} and
\S\ref{pixel}.

\subsection{Test of the anisotropic correction}
\label{ani_test}
\begin{figure}
\psfig{figure=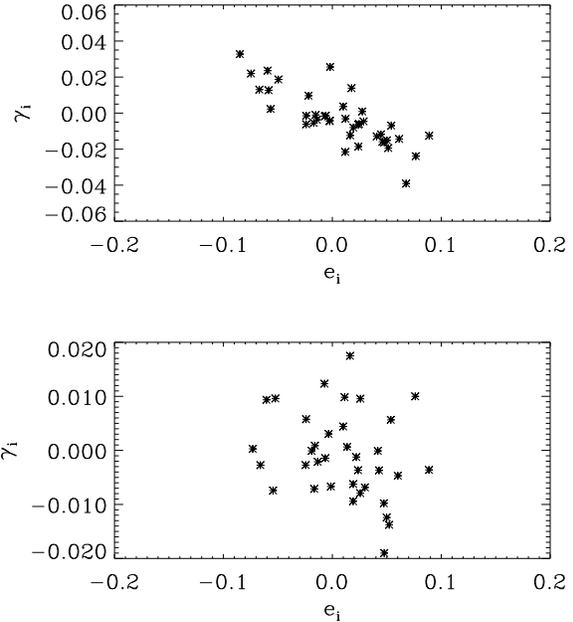,width=80mm} 
\caption{The anti-correlation of
$\bar{e}_i^*$ and $\bar{\gamma}_i$ plotted for 20 simulated fields,
where i=1,2 have been superposed, for (top) a $\nu >5$ cut, and
(bottom) a $\nu >15$ cut. Note the trend for $\nu >5$.}
\label{fig:anti}
\end{figure}

Before we discuss the reclamation of shear by the method, we address
the existence of a remaining systematic effect. In BRE, we found that
a signal/noise cut of $\nu > 5$ (as opposed to our conservative $\nu >
15$) reveals a strong anti-correlation between the mean shear
$\bar{\gamma}_i$ and the mean stellar ellipticity
$\bar{e}^{*}_i$. Here we show that the same effect is found in the
simulated data. Figure \ref{fig:anti} shows $\bar{\gamma}_i$
vs. $\bar{e}^{*}_i$ for 20 simulated fields, which exhibit similar
behaviour to that found in BRE section 6 for real fields. To assess
the significance of this effect, we use the correlation coefficient
\begin{equation}
C_i = \frac {\langle e_i^* \gamma_i \rangle - \langle e_i^* \rangle \langle
\gamma_i \rangle }{\sigma(e_i^*) \sigma(\gamma_i)}.
\end{equation}
For a $\nu > 5$ cut we find $C_1=-0.69$, $C_2=-0.81$ for 20 cells,
which corresponds to a $\gg 3\sigma$ effect as for the real data.  A
cut at $\nu > 15$ reduces the anti-correlation to $C_1=-0.32$,
$C_2=-0.48$. This corresponds to a 1.5-2$\sigma$ effect, which is no
longer a significant contribution to the lensing amplitude. This
anti-correlation is thus due to an over-correction of the PSF for
small galaxies (in Eq.~[\ref{eq:ecorrect}]). This is likely to arise
from the fact that, because of noise, the observed radius (and thus
$(P_{\rm sm}^{g})^{-1}$) of faint galaxies is slightly smaller than
that of the bright stars used to measure the PSF. Note that
figure~\ref{fig:anti} is very similar to the equivalent figure in BRE
section 6 for real data. This again confirms the validity of the
simulations and their use in testing systematic effects, and verifies
that the low-level anti-correlation found in real data stems from a
reproducible problem with the current correction method.

\subsection{Shear Recovery}

We now wish to observe how the output shear derived by the KSB
algorithm compares to the shear input. In order to test this, we ran a
set of simulations with 5\% rms shear for 20 fields. More precisely,
for each field we drew a uniform shear from a Gaussian probability
distribution, with standard deviation equal to a shear of 0.05. A
similar set of 30 fields were simulated with 1.5\% rms shear.

\begin{figure}
\psfig{figure=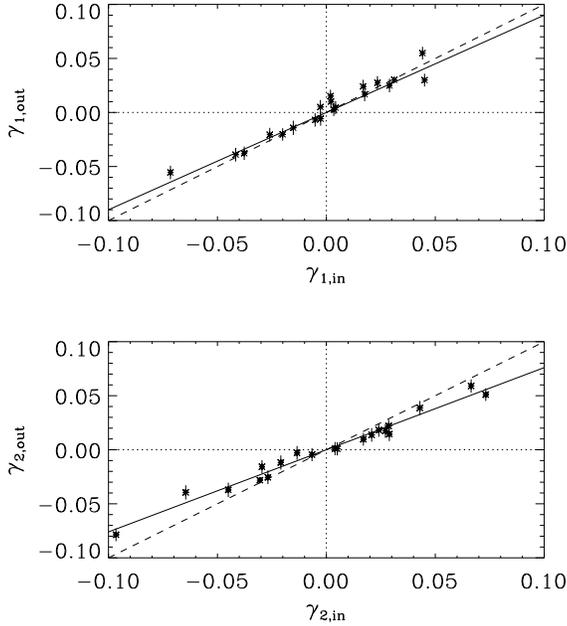,width=80mm}
\caption{$\gamma^{{\rm in}}_{i}$ compared with $\gamma^{{\rm
out}}_{i}$ for simulated data sheared by 5\% rms shear; top panel
shows $\gamma_1$ input and output, while bottom panel shows $\gamma_2$
component. The dashed line shows the $\gamma^{{\rm in}}_{i} =
\gamma^{{\rm out}}_{i}$ relation; the solid lines shows the best fits,
$\gamma^{{\rm in}}_{1} = 0.90 \gamma^{{\rm out}}_{1}$ and
$\gamma^{{\rm in}}_{2} = 0.76 \gamma^{{\rm out}}_{2}$. }
\label{fig:gingout}
\end{figure}

Our results for the 5\% simulations are shown in
figure~\ref{fig:gingout} (see also section 7 in BRE for a
summary). The figure shows that the output shear is clearly linearly
related to the input shear, with a slope close to 1. As a quantitative
test, we apply a linear regression fit to both components of the
shear, combined and separately. For the combined components we obtain
$\gamma^{{\rm out}}_i = 0.0007 + 0.84 \gamma^{{\rm in}}_i$, with standard
errors on the coefficients of 0.001 and 0.04, respectively. For the
individual components we obtain $\gamma^{{\rm out}}_{1} = 0.002 + 0.90
\gamma^{{\rm in}}_{1}$ with errors (.001,.05) and $\gamma^{{\rm
out}}_{2} = 0.0001 + 0.76 \gamma^{{\rm in}}_{2}$ with errors
(.001,.04). For the 1.5\% simulations we similarly obtain consistent
results, namely $\gamma^{{\rm out}}_{i} = 0.0001 + 0.79 \gamma^{{\rm
in}}_{i}$ for combined components with respective standard errors of
0.001 and 0.091.

We see that the {\tt imcat} measure of shear is symmetrical about
zero, but is measuring a slightly smaller shear signal than the input
shear. In similar conditions, we should therefore adjust our shear
measures by dividing $\gamma_1$ by $0.9\pm0.05$ and $\gamma_2$ by
$0.76\pm0.04$ when using this KSB implementation. A full discussion of
the recovery of rms shears using an extensive statistical analysis can
be found in BRE, including discussion of the recovery of rms shears
from sets of simulated fields.

\subsection{Effect of Seeing}
\label{seeing}
Of great practical interest is the dependence of the sensitivity of
weak lensing measurements on seeing. To study this dependence, we ran
several simulations with the same object catalogue, but with different
seeing values, for a set exposure time of 1 hour. For each simulated
$8'\times8'$ simulated field, we computed the rms noise $\sigma_{\rm
noise} \equiv \sigma_\gamma/\sqrt{N}$, where $\sigma_\gamma$ is the
rms of shear measures in a single field, and $N$ is the number of
usable galaxies in the field. The quantity $\sigma_{\rm noise}$ is a
measure of the uncertainty for measuring the average shear in the
field.

\begin{figure}
\psfig{figure=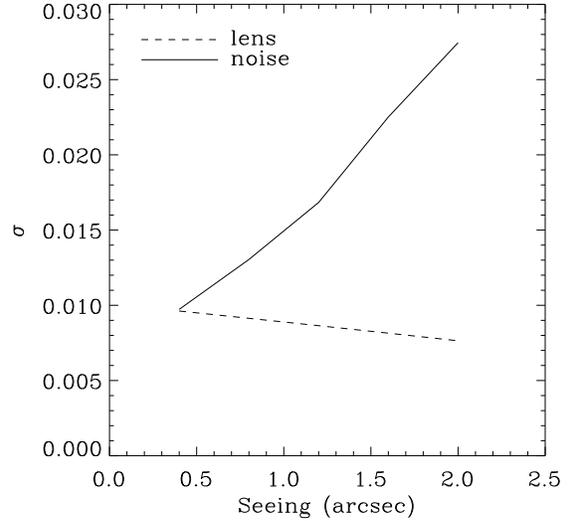,width=80mm}
\caption{$\sigma_{\rm noise}$ (solid) and $\sigma_{\rm lens}$ (dashed)
as a function of seeing FWHM for a set of simulations with 1 hour
integration time. Note the steady degradation of sensitivity with
seeing.}
\label{fig:seeing}
\end{figure}

The results are shown in table~\ref{tab:seeing} and
figure~\ref{fig:seeing}. As can be seen in the figure, the seeing
degrades the uncertainty almost linearly. Interestingly, the loss of
sensitivity comes primarily from the loss in the number $N$ of usable
galaxies, with no strong increase observed in $\sigma_{\gamma}$
(see table~\ref{tab:seeing}). One might suppose that this degradation
could be countered by longer integrations on the field, to regain the
number counts diluted by the larger isotropic smear. However, besides
the integration time increase being considerable for such a
reclamation of number density (see section \ref{time}), many of the
regained galaxies will still need to be excluded as their shape
information has been erased by a kernel significantly larger than
their intrinsic radius. The noise could perhaps be reduced by improved
shear-measurement methods, which reduce the cuts we have to make on
small galaxies.

Note that, for worse seeing cases, the usable galaxies will be on
average brighter and larger and will thus have a lower median
redshift. This will tend to degrade the lensing signal further.  For a
cluster normalised CDM model, the shear rms from lensing in an
8'$\times$8' cell is $\sigma_{\rm lens} \simeq 0.012 z_{m}^{0.8}$
(BRE). The median redshift $z_{m}$ of the galaxies is derived from the
median $R$-magnitude using the results of Cohen et al (2000).  The
resulting lensing rms $\sigma_{\rm lens}$ is also plotted as a
function of seeing in figure~\ref{fig:seeing}, and the signal-to-noise
ratio for a single (8'x8') cell, $S/N = \sigma_{\rm lens}/\sigma_{\rm
noise}$ is listed in table \ref{tab:seeing}. We find that the
reduction of $\sigma_{\rm lens}$ with poorer seeing is rather
weak. Thus the reduction of signal-to-noise for shear measurement is
again dominated by the decrease in $N$.

\begin{table*}
\centering \begin{minipage}{180mm}
\caption{Shear sensitivity as a function of seeing, in (8'$\times$8') cells.}
\label{tab:seeing}
\begin{tabular}{llllllll} \hline
Seeing ('') & $n_g$ (arcmin$^{-2}$) & $\sigma_\gamma$ & $\sigma_{\rm noise}$ &
Median R & Median z & $\sigma_{\rm lens}$ & S/N\\
\hline
0.4 & 29.8 & 0.43 & 0.0097 & 24.1 & 0.8 & 0.0096 & 1.0\\
0.8 & 18.0 & 0.44 & 0.0130 & 23.4 & 0.8 & 0.0096 & 0.7\\
1.2 & 11.8 & 0.46 & 0.0168 & 22.8 & 0.7 & 0.0086 & 0.5\\ 
1.6 &  7.5 & 0.49 & 0.0225 & 22.2 & 0.6 & 0.0076 & 0.3\\ 
2.0 &  5.2 & 0.50 & 0.0275 & 21.9 & 0.6 & 0.0076 & 0.3\\
\hline
\end{tabular}\\
\end{minipage}
\end{table*}

\subsection{Effect of Integration Time}
\label{time}
\begin{figure}
\psfig{figure=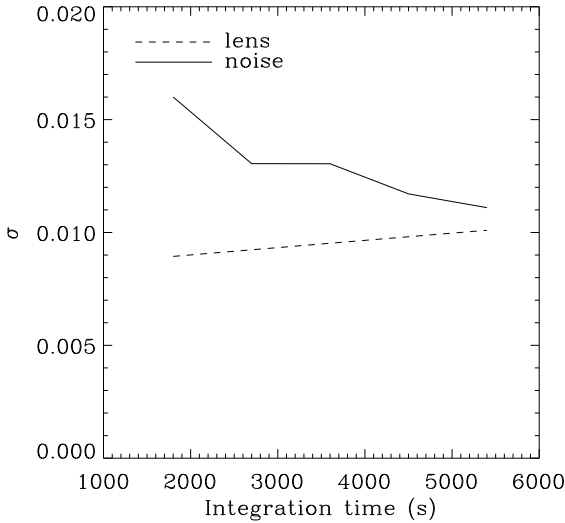,width=80mm}
\caption{Effect of integration time on the noise (solid) and lensing
(dashed) rms.}
\label{fig:time}
\end{figure}

To optimise weak lensing surveys, one needs to compromise between
depth and width. To help in this optimisation, we produced several
simulated images for different exposure times, while keeping the
seeing at 0.8''. Table~\ref{tab:time} shows the quantities discussed
in the previous section for different exposure times relevant for
ground-based observations. The noise and lensing rms are plotted on
figure~\ref{fig:time}. The dependence of these quantities on exposure
time is rather weak. This is due to the fact that the fainter galaxies
which can be detected with deeper exposures are too small to be
resolved in the presence of seeing, and must therefore be mostly
discarded. Moreover, since intrinsic ellipticities dominate the total
ellipticity dispersion in this regime (see \S\ref{e_dist}), the
reduced pixel noise of deeper images does not substantially reduce
$\sigma_{\rm noise}$. As a result, the signal-to-noise $S/N$ to
measure lensing is only moderately improved for longer exposures.

\begin{table*}
\centering \begin{minipage}{180mm}
\caption{Shear sensitivity as a function of integration time, for
(8'$\times$8') cells.}
\label{tab:time}
\begin{tabular}{llllllll} \hline
time (s) & $n_g$ (arcmin$^{-1}$) & $\sigma_\gamma$ & $\sigma_{\rm noise}$ &
Median R & Median z & $\sigma_{\rm lens}$ & S/N \\
\hline
1800 & 11.3 & 0.43 & 0.0160 & 22.9 & 0.7 & 0.0086 & 0.5\\
2700 & 16.4 & 0.42 & 0.0131 & 23.4 & 0.8 & 0.0096 & 0.7\\
3600 & 18.0 & 0.44 & 0.0130 & 23.4 & 0.8 & 0.0096 & 0.7\\
4500 & 20.4 & 0.42 & 0.0117 & 23.6 & 0.8 & 0.0096 & 0.8\\
5400 & 22.0 & 0.42 & 0.0111 & 23.8 & 0.9 & 0.0106 & 1.0\\
\hline
\end{tabular}\\
\end{minipage}
\end{table*}

\subsection{Effect of Pixelisation}
\label{pixel}
Another test of practical interest is the dependence of the
sensitivity to lensing on pixel size. This is important for the design
of future dedicated instruments (eg. Kaiser et al 2000, Tyson et al
2000). To study this dependence in the context of current, ground-based
observations, we again produced several simulated fields from the same
object catalog, keeping a seeing FWHM of 0.6'' and exposure time of 1
hour, but with different pixel sizes, ranging from 0.1'' to 1.0''.

\begin{figure}
\psfig{figure=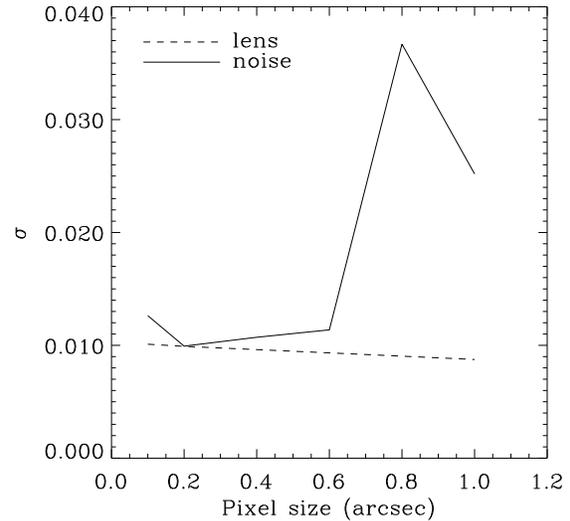,width=80mm}
\caption{Effect of pixel size on the noise (solid) and lensing
(dashed) rms.}
\label{fig:pixel}
\end{figure}

The results are listed in Table~\ref{tab:pixel} and
Figure~\ref{fig:pixel}. We find that the noise is quite stable for
pixel scales smaller than the seeing radius. Increased oversampling of
the PSF does not improve the noise properties to any great
degree. However, as the pixel scale increases above the seeing FWHM,
the method fails quickly. The stellar locus on a magnitude-radius plot
then approaches the galaxy locus even at bright magnitudes, making
star selection for anisotropic smear correction very
difficult. Moreover, a pixel scale of, say, 0.8'' with a seeing FWHM
of 0.6'' are both conspiring together to remove shape information
grievously from galaxies beyond $R \simeq 22$. As a result,
$\sigma_{\rm noise}$ rapidly grows. Extreme oversampling of the PSF
appears to be inefficient, while under-sampling is very detrimental for
typical ground-based seeing. Under-sampling is less of a problem for
space-based data, however, since the typical pixel scale and PSF FWHM
(say 0.1'') are much smaller than the typical galaxy radius at
magnitudes of interest (eg. Rhodes, Refregier \& Groth 2000).

\begin{table*}
\centering \begin{minipage}{180mm}
\caption{Shear sensitivity as a function of pixel size, for (8'$\times$8')
cells}
\label{tab:pixel}
\begin{tabular}{llllllll} \hline
Pixel ('') & $n_g$ (arcmin$^{-1}$) & $\sigma_\gamma$ & $\sigma_{\rm noise}$ &
Median R & Median z & $\sigma_{\rm lens}$ & S/N \\
\hline
0.1 & 23.9 & 0.43 & 0.0126 & 23.5 & 0.8 & 0.0096 & 0.8\\
0.2 & 23.4 & 0.33 & 0.0099 & 23.5 & 0.8 & 0.0096 & 1.0\\
0.4 & 21.7 & 0.35 & 0.0107 & 23.4 & 0.8 & 0.0096 & 0.9\\
0.6 & 19.5 & 0.35 & 0.0114 & 23.4 & 0.8 & 0.0096 & 0.8\\
0.8 & 12.1 & 1.02 & 0.0367 & 23.1 & 0.7 & 0.0086 & 0.2\\
1.0 & 13.7 & 0.66 & 0.0252 & 23.0 & 0.7 & 0.0086 & 0.3\\
\hline
\end{tabular}\\
\end{minipage}
\end{table*}

\subsection{Effect of Overlapping Isophotes}
Spurious lensing signals could also be produced on small scales by
overlapping isophotes of neighboring galaxies. Van Waerbeke et al.
(2000) suggested that this effect could explain the excess small scale
power observed in their cosmic shear surveys. To understand this
effect, consider two galaxies which, for simplicity, are
assumed to be circular. If the galaxies are separated by an angular distance
comparable to their angular sizes, their combined isophotes will
have a dumb-bell shape rather than be the simple sum of two disjoint
circular isophotes. As a result, their ellipticity will tend to be
aligned along the separation axis, leading to a spurious ellipticity
correlation.

To test the impact of this effect, we performed 10 simulations without
lensing shear (with seeing 0.8'' and exposure time of 1 hour). For these
simulations, we measured the shear correlation functions defined as
(eg. Kamionkowski et al.  1998)
\begin{equation}
C_{ij}(\theta) \equiv \langle \gamma_{i}(0) \gamma_{j}(\theta) 
  \rangle,
\end{equation}
where $i$ and $j$ run from 1 to 2. Here, $\gamma_{i}(0)$ and
$\gamma_{j}(\theta)$ are the shear estimates of each member of a
galaxy pair with a separation $\theta$. These ellipticities are
measured in a coordinate system whose x-axis is along the separation
angle of the galaxy pair (see Heavens, Refregier \& Heymans (2000) for
an illustration). After measuring these correlation functions for each
of the 10 simulations separately, we computed the mean and error in
the mean over all simulations.

\begin{figure}
\psfig{figure=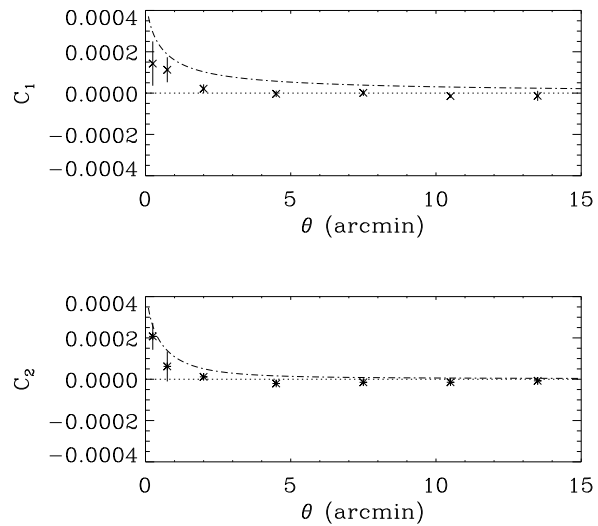,width=80mm}
\caption{Mean correlation functions $C_1$ and $C_2$ for 10 null
simulation fields. Points show the mean of 10 correlation functions,
with associated uncertainty. The dash-dotted curves are the
correlation functions for the expected lensing signal for a
cluster-normalised $\Lambda$CDM cosmology.}
\label{fig:cfn}
\end{figure}

The resulting correlation functions are shown on Figure
\ref{fig:cfn}. As a comparison, the correlation function expected from
lensing for a cluster-normalised $\Lambda CDM$ model is also shown
(see ibid for details of the calculation).  In the null simulation, we
weakly detect a correlation on scales smaller than about 1'. The
amplitude of the overlapping isophote effect is $C_{i} \simeq 10^{-4}$
corresponding to an rms shear of about 1\%. As can be seen on the
figure, this is smaller, but comparable to the lensing signal expected
on these scales. The exact amplitude of the overlapping isophote
effect will depend on the precise conditions of the observation (or
simulation). It is nevertheless likely that the excess power observed
by van Waerbeke et al. (2000) on small scales ($\theta \la 10''$) is
indeed due to this effect. Apart from residual systematic effects,
another explanation for this excess power could be the intrinsic
alignment of galaxies. Theoretical studies however indicate that this
effect is small for a survey of this depth (Heavens, Refregier \&
Heymans 2000; Metzler \& Croft 2000; see also Catelan, Kamionkowski \&
Blandford 2000, and Pen, Lee \& Seljak 2000).

\section[]{Conclusions}
\label{conclusion} 
We have tested our shear measurement method using numerical
simulations of artificial images. The object catalogues were created
by generating realisations of the HST Groth Strip; the resulting
artificial images include observational effects such as noise, seeing,
and anisotropic PSF.  We compare our realised catalogues to those
observed with WHT and find good statistical agreement. We used these
simulations to test the KSB shear measurement method. Overall, we find
that this method is rather accurate, but with several provisos: we
find a residual anti-correlation between the PSF ellipticity and the
corrected ellipticities of faint galaxies. This effect can be made
negligible if faint galaxies (with $S/N \la 15$) are removed from the
catalogue. We also find that the recovered shear is linearly related
to the input shear, but with a coefficient of about 0.8 which must be
used to calibrate the final shear. With these precautions, the KSB
method is sufficient for the current weak lensing surveys. However,
the method is neither optimal nor necessarily extendable to superior
observing conditions. It should therefore be replaced with more
accurate methods such as that of Kaiser (1999b), Rhodes, Refregier \&
Groth (2000), and Kuijken (1999), in future, more sensitive surveys.

We also used our simulations to study the effect of seeing, exposure
time and pixelisation on the sensitivity to the shear. We found that
increased seeing FWHM increases the noise almost linearly, with the
primary loss being the decreased number of usable galaxies. In the
seeing-dominated regime, the sensitivity to shear is only weakly
dependent on exposure time. As long as this regime holds, it is
therefore more efficient to tend towards wide rather than deep weak
lensing surveys. Increased pixel scale hardly affects the sensitivity
until the pixel scale is comparable to the seeing FWHM, at which point
the method fails for typical ground-based seeing. Thus, extreme
oversampling of the PSF does not seem to be necessary.

We also tested the claim by van Waerbeke et al. that spurious shear
signals on small scales ($\theta \la 10''$) could be produced by
overlapping isophotes of neighboring galaxies. Using simulated images
without input shear, we weakly detect this effect on scales $\theta
\la 1'$. The rms amplitude of this effect is of about 1\%, which is
smaller but comparable to that expected for lensing. Overlapping
isophotes are thus likely to explain the excess power found
on small scale by this group.

\section*{Acknowledgements}
We would like to thank Thomas Erben, Peter Schneider, Yannick Mellier,
Roberto Maoli and Aur\'elien Thion for useful discussions. We are
indebted to Nick Kaiser for providing us with the Imcat software. We
acknowledge the invaluable use of IRAF and SExtractor during this
research. AR was supported by a TMR postdoctoral fellowship from the
EEC Lensing Network, and by a Wolfson College Research Fellowship. DC
acknowledges the ``Sonderforschungsbereich 375-95 f\"ur
Astro--Teil\-chen\-phy\-sik" der Deutschen
For\-schungs\-ge\-mein\-schaft for financial support.  This work was
supported by the TMR Network ``Gravitational Lensing: New Constraints
on Cosmology and the Distribution of Dark Matter'' of the EC under
contract No. ERBFMRX-CT97-0172.

\bsp

\label{lastpage}

\end{document}